\documentclass[apj]{emulateapj}
\usepackage{natbib}
\usepackage[caption=false]{subfig}
\usepackage{graphicx}
\usepackage{makecell}
\shorttitle{DCBH formation}
\shortauthors{Dunn et al.}
\usepackage{color}

\begin{document}

\title{Sowing black hole seeds: Direct Collapse Black Hole Formation with realistic Lyman-Werner Radiation in cosmological simulations}

\author{Glenna Dunn\altaffilmark{1}, Jillian Bellovary\altaffilmark{2,3}, Kelly Holley-Bockelmann\altaffilmark{1,4}, Charlotte Christensen\altaffilmark{5}, \and Thomas Quinn\altaffilmark{6}}
\affil{1 Vanderbilt University, Nashville, TN, USA}
\affil{2 Queensborough Community College, New York, NY, USA}
\affil{3 American Museum of Natural History, New York, NY, USA}
\affil{4 Fisk University, Nashville, TN, USA}
\affil{5 Grinnell College, Grinnell, IA, USA}
\affil{5 University of Washington, Seattle, WA, USA}
\email{glenna.dunn@vanderbilt.edu}

\begin{abstract}
We study the birth of supermassive black holes from the direct collapse process and characterize the sites where these black hole seeds form.  In the pre-reionization epoch, molecular hydrogen (H$_2$) is an efficient coolant, causing gas to fragment and form Population III stars, but Lyman-Werner radiation can suppress H$_2$ formation and allow gas to collapse directly into a massive black hole. The critical flux required to inhibit H$_2$ formation, $J_{\rm crit}$, is hotly debated, largely due to the uncertainties in the source radiation spectrum, H$_2$ self-shielding, and collisional dissociation rates. Here, we test the power of the direct collapse model in a self-consistent, time-dependant, non-uniform Lyman-Werner radiation field -- the first time such has been done in a cosmological volume -- using an updated version of the SPH+N-body tree code {\sc Gasoline} with H$_2$ non-equilibrium abundance tracking, H$_2$ cooling, and a modern SPH implementation. We vary $J_{\rm crit} $ from $30$ to $10^3$ in units of $J_{21}$ to study how this parameter impacts the number of seed black holes and the type of galaxies which host them. We focus on black hole formation as a function of environment, halo mass, metallicity, and proximity of the Lyman-Werner source.  Massive black hole seeds form more abundantly with lower $J_{\rm crit}$ thresholds, but regardless of $J_{\rm crit}$, these seeds typically form in halos that have recently begun star formation.  Our results do not confirm the proposed atomic cooling halo pair scenario; rather black hole seeds predominantly form in low-metallicity pockets of halos which already host star formation.
\end{abstract}


%%%%%%%%%%%%%%%%%%%%%%
\section{Introduction}
%%%%%%%%%%%%%%%%%%%%%%
The detections of luminous quasars at redshifts $z\gtrsim 6$ \citep[e.g.][]{Fan03,Kurk07,Mortlock11,DeRosa14,Banados16} indicate that the first black holes formed early and grew to $10^{6}-10^{9}M_{\odot}$ within the first billion years after the Big Bang.  While the formation mechanism and early growth of these massive black holes (MBH) is widely debated, these objects likely formed from massive `seed' black holes \citep{Haiman01} within the first few hundred million years after the Big Bang and accumulated mass swiftly to produce the observable distribution of high-redshift quasars \citep{Volonteri03}.  The evolution of an MBH and its host halo are integrally tied, and this relationship can be harnesses to explore heirarchical structure formation \citep[e.g.][]{Micic07} and probe MBH demographics \citep[e.g.][]{Micic08,Volonteri09}.  

One popular model of MBH formation proposes that MBH seeds with masses of ~$100 M_{\odot}$ form from the remnants of high-mass population III stars \citep{Madau01, Heger02, Johnson07}(see, however, \citet{Stacy10,Hosokawa11}).  Such low-mass seeds may serve as building blocks for supermassive black holes through both MBH-MBH mergers and gas accretion \citep[e.g.][]{Micic07,HolleyBockelmann10}. A second possible formation channel may be through stellar collisions in a dense nuclear star cluster \citep{Begelman78,Devecchi09,Davies11,Katz15,Yajima16}.  A third theory, which we study in this paper, is that MBHs form by the direct isothermal collapse of pristine gas within a primordial dark matter halo \citep{Loeb94,Haiman96,Koushiappas04,Begelman06,Lodato06,Haiman06,Spaans06,Shang10,Prieto13,Latif13b,Latif13c,Johnson14,Choi15,Glover16,Yue17}.  Because the seeds are massive in this model, direct collapse has been invoked to explain the massive quasars at high redshift \citep{Natarajan17}.  In this work, we study the demographics of direct collapse black hole formation sites in the context of fully cosmological simulations with a self-consistent radiation field. 

In this paper, we consider thermodynamic direct collapse enabled by Lyman-Werner radiation.  Alternative mechanisms that may suppress fragmentation could allow early dark matter halos to remain dynamically hot and support the gas against collapse despite efficient thermal cooling.  Baryonic streaming velocities, left over from baryon-photon fluid decoupling at recombination, increase the random motion of gas in high-redshift halos and may delay collapse long enough to allow these halos to acquire enough mass to enable preferential formation of MBH seeds  \citep{Schauer17a,Tanaka14,Hirano17}.  Dynamical heating effects due to the rapid growth of halos through accretion and mergers can also foster random motion of gas in high-redshift halos and may delay collapse, effectively increasing the final mass of the collapsing object \citep{Yoshida03}.

Conditions that would permit thermodynamic direct collapse to proceed without fragmentation are likely rare, as it requires that the Jeans mass of the collapsing cloud remain large.  This environment is expected to only occur in pristine atomic cooling halos with $T_{vir}>10^4$K, corresponding to a redshift dependent minimum halo mass of $M_{halo}>3\times 10^{7}M_{\odot}[(1+z)/11]^{-3/2}$ \citep{Visbal17,Barkana01}.  Furthermore, these halos must have metallicity $\rm Z/Z_{\odot} \leq 5\times 10^{-6} - 10^{-5}$ \citep{Omukai08,Clark08,Latif16} and experience intense Lyman-Werner radiation from a nearby starforming region \citep{Latif13c,Regan14}.  These constraints imply a small window of opportunity for MBH formation in cosmic time, halo mass, and location.

If gas cools too quickly, it easily fragments; in essence, the direct collapse model relies on removing efficient gas cooling channels.  Before the first stars enriched their host halos with metals, the primary coolant for halos below a virial temperature of $\sim8,000$K is molecular Hydrogen (H$_2$).  In this scenario, the gas must remain pristine not only because metal-line cooling enhances fragmentation, but because Hydrogen molecules H$_2$ and HD form readily on dust grains \citep{Cazaux09}, adding another avenue for cooling to proceed efficiently.  Without H$_2$ or metals, gas in these proto-galactic halos can only cool though collisional excitation of atomic Hydrogen.  Atomic hydrogen cannot cool the gas below T$\sim10^4$ K, resulting in a Jeans mass of $10^4-10^5 M_{\odot}$.  To clear a halo of H$_2$, its gas must be irradiated with UV photons to photo-dissociate any molecular Hydrogen within; naturally the energy range to disrupt H$_2$ corresponds to the Lyman and Werner absorption bands between 11.2 - 13.6 eV.  This specific intensity necessary to suppress molecular Hydrogen in atomic cooling halos is known as $J_{\rm crit}$, and is expressed in units of $J_{21}$ ($10^{-21}$ erg s$^{-1}$ cm$^{-2}$ $sr^{-1}$ Hz$^{-1}$).

In summary, a halo with a virial temperature $T_{\rm vir}\ge 10^4 K$ exposed to Lyman-Werner radiation in excess of a critical threshold, $J_{\rm crit}$, can collapse isothermally under its own gravity directly into a black hole, without fragmenting to form stars \citep{Shang10,Latif14}.  The critical value of Lyman-Werner radiation required to prevent $H_{2}$ formation in a halo, $J_{\rm crit}$, is unknown, and possible values presently span several orders of magnitude, reaching as low as $30$ if produced by a soft stellar spectrum or as high as $10^4$ from a hard stellar spectrum \citep{Shang10, Glover15b}.  Complicating the simple picture of a single radiation threshold, \citet{Sugimura14}, \citet{Agarwal16} and \citet{WolcottGreen17} argue that in fact there may be no universal $J_{\rm crit}$, but instead that the gas density, as well as the mass, age and distance to the Lyman-Werner source all generate variable Lyman-Werner irradiation that ultimately manifests in halo-to-halo variations in $J_{\rm crit}$.

Following the lead of theoretical and semi-analytic studies \citep{Dijkstra08,Ahn09,Dijkstra14}, the bulk of numerical work has focused on treatments in which seed MBHs form according to prescribed global halo properties.  \citet{Agarwal14} restricted their studies of direct collapse formation sites to halos that are strictly metal free, with no star formation history, and have only been exposed to Lyman-Werner radiation from non-local sources.  \citet{Habouzit16} used several different volume cosmological simulations to identify direct collapse seed locations in metal-poor atomic cooling halos testing three different $J_{\rm crit}$ values and two different formation timescales. \citet{Tremmel17} used state-of-the-art cosmological simulations constructed with ChaNGa to model MBH seed formation in metal-poor, rapidly collapsing, slowly cooling gas particles, but does not consider the role of Lyman-Werner radiation or H$_2$ cooling.  The black holes in these simulation grow nearly instantaneously to $10^6$ M$_{\odot}$.  Several of the modern interpretations of the direct collapse model invoke the synchronization of two neighboring atomic cooling halos separated by less than 1-2 kpc.  One halo experiences a burst of star formation that provides enough Lyman-Werner radiation to prevent star formation in the neighboring halo \citep[see e.g.][]{Dijkstra08,Shang10,Visbal14a,Regan14,Regan17,WolcottGreen17}.

In this work, we study the formation of direct collapse MBH seeds as governed by a self-consistently modeled, non-uniform Lyman-Werner radiation field in a cosmological volume.  This seeding method makes no assumptions about the global properties of the halos that form MBHs, nor does it make any requirements of the halo's history.  Instead, MBH seeds form from gas particles according to the local gas properties and Lyman-Werner radiation levels.  This allows us the unique opportunity to study the direct collapse model in a fully self-consistent environment and extract halo properties of MBH hosts in post-processing.  This work is the first ever treatment of massive black hole formation based solely upon local gas physics in a cosmological context to include a self-consistent spatially and temporally varying Lyman-Werner radiation field.

The paper is organized as follows: in Section \ref{Sims} we discuss the simulation suite; in Section \ref{LW} we discuss the role of Lyman-Werner radiation in direct collapse; in Section \ref{MBH} we outline our MBH seed formation algorithm; and in Section \ref{Results} we discuss the implications of low-$J_{\rm crit}$ MBH formation models for low-redshift MBH distributions.

%%%%%%%%%%%%%%%%%%%%%%%%%%%%%%%%%%
\section{Simulations} \label{Sims}
%%%%%%%%%%%%%%%%%%%%%%%%%%%%%%%%%%
In this work, we use the N-body+Smooth Particle Hydrodynamics (SPH) tree code \textsc{Gasoline} \citep{Stadel01,Wadsley04,Wadsley17} to study the formation of MBH seeds in cosmological simulations.  \textsc{Gasoline} has previously been shown to form galaxies that follow the Tully-Fisher relation \citep{Governato09}, the halo mass-metallicity relationship \citep{Brooks07}, the stellar mass $-$ halo mass relation \citep{Munshi13}, size-luminosity relation \citep{Brooks11}, and produce realistic disks, bulges, and star formation histories \citep{Christensen10,Christensen14a}.  

This implementation of \textsc{Gasoline} includes a geometric density average in the SPH force expression \citep{Ritchie01}.  The treatment of gas physics includes an H$_{2}$ non-equilibrium abundance tracking and cooling procedure, a sophisticated treatment of H$_{2}$ abundance calculations that account for self-shielding, collisional dissociation, and dust grain formation and shielding \citep{Christensen12}.  We also consider the photo-dissociation rate due to spatially and temporally varying Lyman-Werner flux produced self-consistently by stars in the simulation, \citep{Christensen12} (see Section \ref{MBHcode} for a more complete discussion of the implementation and role of Lyman-Werner radiation in these simulations).  In addition to H$_{2}$ cooling, the simulation includes metal-line cooling and turbulent metal diffusion \citep{Shen10}.  The simulation includes a uniform background ionizing radiation beginning at a redshift of $z=9$ \citep{Haardt96}, which is independent of the Lyman-Werner radiation generated by stars. The simulation also includes the Sedov solution blast-wave supernova feedback energy with a blast-wave energy of $E_{SN}=8\times 10^{50}$ erg \citep{Stinson06}.  Star formation occurs stochastically according to a Kroupa stellar initial mass function \citep{Kroupa01}, with a probability that depends on the H$_2$ fraction of the gas.  \textsc{Gasoline} also includes physically motivated prescriptions for black hole formation, accretion, feedback and mergers \citep{Bellovary10}.  For the first time, we include a revised black hole formation prescription featuring a probabilistic function that depends on the local Lyman-Werner flux, $J_{LW}$.  We discuss the implementation of star formation and MBH seed formation physics in \textsc{Gasoline} and our modifications to the black hole formation model are discussed in Section \ref{MBH}. 

We tested three different values of $J_{\rm crit}$, spanning the range of predicted values ($J_{\rm crit}=10^3 J_{21}$ $300 J_{21}$, and $30 J_{21}$, see Section \ref{MBH}) by examining black hole formation in a Milky Way-analog disk galaxy ($h258$).  The evolution of this halo has previously been studied in detail \citep[see e.g.][]{Bellovary10,Bellovary11,Zolotov12,Munshi14,Christensen16}.  The initial conditions for this galaxy were selected from a 50 Mpc uniform dark matter-only simulation that was re-simulated at a higher resolution via volume renormalization \citep{Katz93}.  At $z=5$, the main progenitor of this halo has reached a mass of $8.5\times 10^{9} M_{\odot}$.  Each of the simulations has a dark matter mass resolution of $3.7 \times 10^4 M_{\odot}$, a gas mass resolution of $2.7 \times 10^4 M_{\odot}$, a stellar mass resolution of $8\times 10^3 M_{\odot}$ and a spline kernel softening length of $0.173$ kpc.  Since the focus of this work is the era of MBH formation, we evolve our simulations from the initial redshift of $z=149$ to $z=5$.

\begin{figure*}[ht]
  \centering
	\subfloat{\includegraphics[width=\columnwidth]{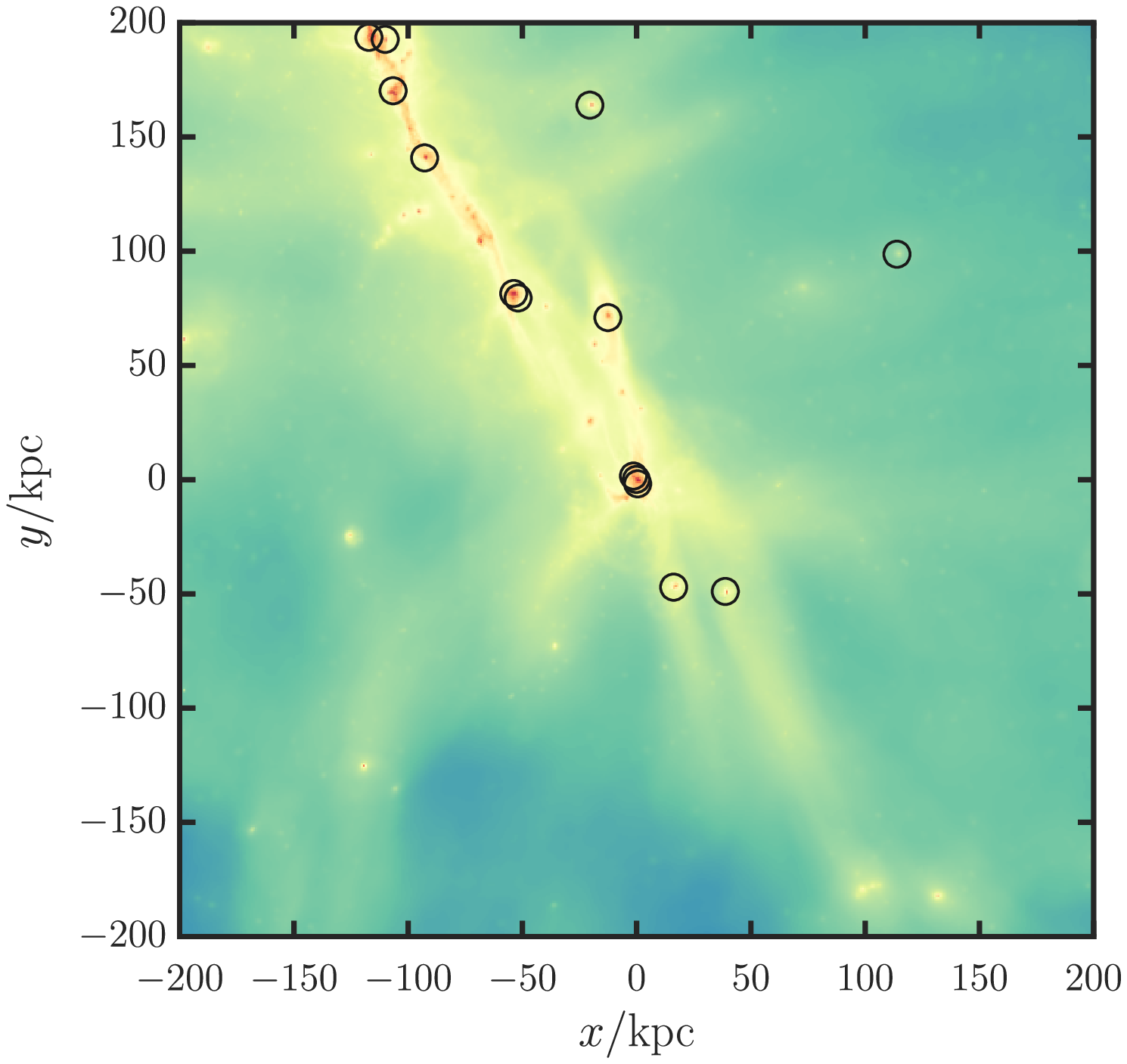}}
	\subfloat{\includegraphics[width=\columnwidth]{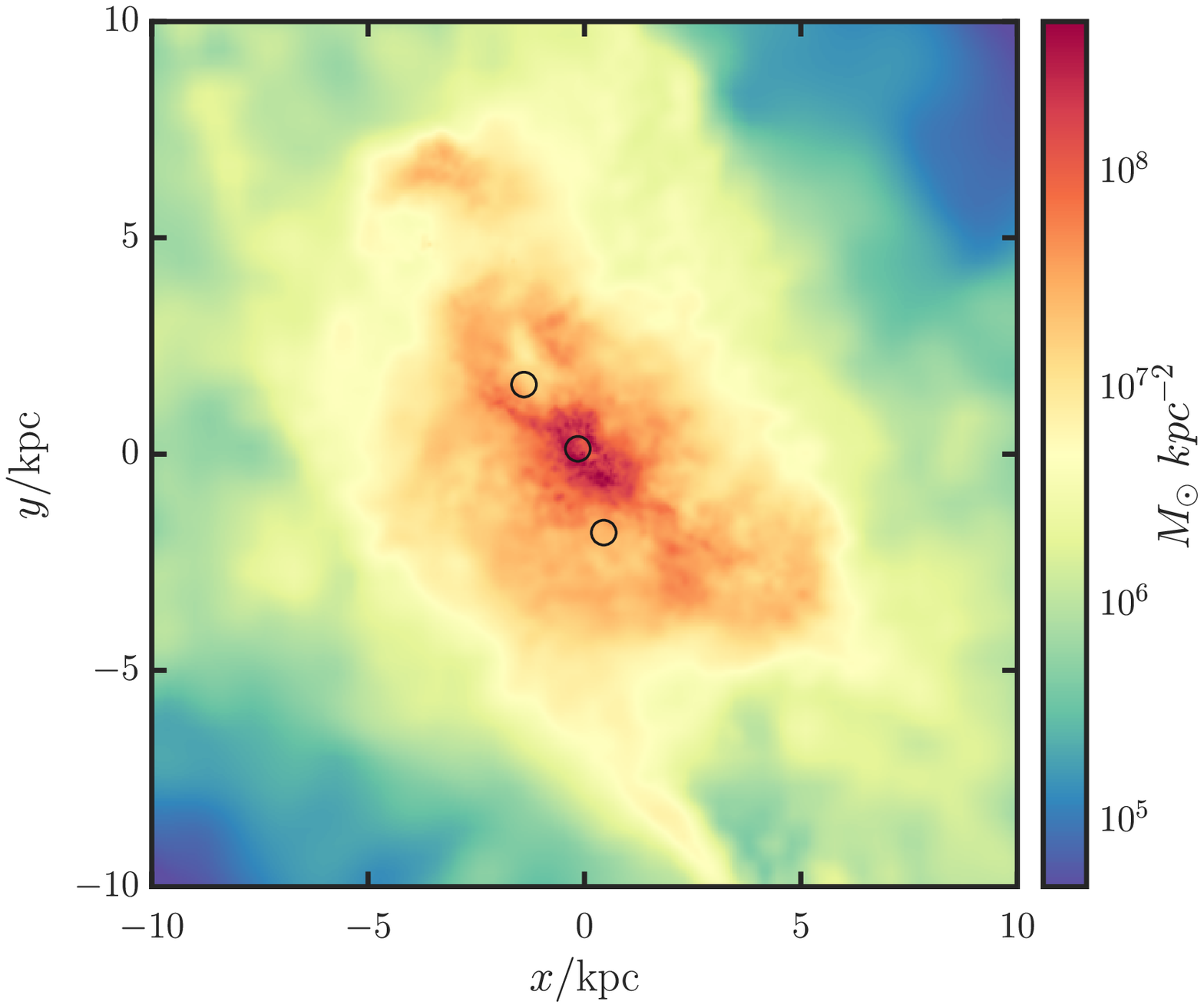}}
	\caption{\small{\textsc{mbh formation sites at redshift 5} Left: Massive black hole seed positions are marked by black circles in this $z=5$ gas surface density image of the $h258$ progenitor halos constructed from the $J1000$ simulation.  The image is centered on the most massive progenitor halo at this redshift, which hosts three MBH seeds.  Right: An enlarged image of the most massive progenitor halo at $z=5$ shows that this $2.7\times 10^{10} M_{\odot}$ protogalaxy hosts three centrally located MBHs, the most massive of which is $2.7\times 10^5 M_{\odot}$.}}
  	\label{h258redshift5}
\end{figure*}

The simulations are fully cosmological, initialized at $z=149$ with the Zel'dovich approximation and WMAP3 cosmology: $\Omega_{m}=0.24$, $\Omega_{b}=0.042$, $\Omega_{\Lambda}=0.76$, $h=0.73$, $\sigma_{8}=0.77$, and $n=0.96$ \citep{Spergel07}.  We note here that previous work has shown that halos in cosmological simulations initialized with the Zel'dovich approximation form later, leading to later black hole formation, than in simulations initialized with second-order Lagrangian perturbation theory \citep{HolleyBockelmann12}.  We identified halos using the Amiga Halo Finder \citep{Gill04,Knollmann09}, which employs isodensity surfaces to identify gravitationally bound objects, to identify halos with more than 64 particles.  Halo centers are identified using the shrinking-sphere method of \citet{Power03} as implemented in Pynbody \citep{Pontzen13}.  We consider halos with more than $10^3$ particles, corresponding to a minimum mass of $\sim 10^7 M_{\odot}$, well resolved.  We note here that this resolution allows us to resolve most atomic cooling halos, which would have a minimum mass of $\sim 2-3\times 10^{7}M_{\odot}$ during the period of peak MBH formation.  In Figure~\ref{h258redshift5}, we show a line-of-sight averaged image of the gas, colored by density, in the high-resolution region of the $J_{crit}=10^3J_{21}$ simulation at $z=5$.  The locations of MBH seeds are marked by open black circles.  At the end of the simulation, there are 15 MBHs in the $J1000$ simulation, three of which reside in the most massive halo.  Several halos host more that one MBH by $z=5$.

%%%%%%%%%%%%%%%%%%%%%%%%%%%%%%%%%%%%%%%%%%%%%%%%%%%%%%
\section{MBH seed formation and evolution} \label{MBH}
%%%%%%%%%%%%%%%%%%%%%%%%%%%%%%%%%%%%%%%%%%%%%%%%%%%%%%%%%%%%%%%%%%%%%%%%%%%%
\subsection{Lyman-Werner radiation and the atomic cooling regime} \label{LW}
%%%%%%%%%%%%%%%%%%%%%%%%%%%%%%%%%%%%%%%%%%%%%%%%%%%%%%%%%%%%%%%%%%%%%%%%%%%%
The current range of possible $J_{\rm crit}$ values spans several orders of magnitude \citep{Haiman00,Latif13a,Aykutalp14,Latif15}.  The uncertainty in this value stems in part from the spectral shape of the radiation sources providing the Lyman-Werner radiation \citep{Latif14}, the H$_{2}$ self-shielding model \citep{WolcottGreen11}, and the gas chemistry model \citep{Glover15a}.  Beyond these model-dependent uncertainties, it is possible that there is no universal $J_{\rm crit}$ value \citep{Agarwal16} but instead that there is critical relationship between the rate of photo-dissociation of H$_2$ molecules by Lyman-Werner photons and the rate of photo-detachment of H$^{-}$ by infrared light that governs whether the H$_2$ fraction rises high enough to enable catastrophic cooling and fragmentation \citep{WolcottGreen17}.  Even so, it is clear that some minimum level of Lyman-Werner radiation is necessary to fully suppress molecular Hydrogen cooling; if the incident flux is lower than $\sim10 J_{21}$, fragmentation is not prevented but only delayed, and may ultimately yield increased fragmentation \citep{Regan18}.

The direct collapse model of MBH seed formation requires that $10^4-10^6 M_{\odot}$ of gas loses enough angular momentum to rapidly collapse, with gas inflow rates $\ge 0.1 M_{\odot}/\rm yr$ for $\sim1-10$ Myr \citep{Hosokawa13,Alexander14,Umeda16}.  The dissipation of large amounts of angular momentum has been a major theoretical hurdle for the direct collapse model, and has been the subject of much work.  While some authors have restricted the direct collapse domain to only low angular momentum halos \citep{Eisenstein95} or material in the low-angular momentum tail of the specific angular momentum distribution in dark matter halos \citep{Koushiappas04}, other authors have explored mechanisms of angular momentum transport.  Work by \citet{Dubois12} has shown that a significant fraction of the gas entering a high-$z$ halo can stream directly to the central bulge through efficient angular momentum re-distribution.  Such large accretion rates allow the gas to collapse without fragmenting to form stars and can occur through dynamical processes, such as the `bars-within-bars' gravitational instability \citep{Shlosman89,Begelman06,Lodato06}, torques, and turbulence \citep{Choi13,Choi15}.  Furthermore, these angular momentum transport mechanisms can actually serve to further suppress gas fragmentation \citep{Begelman09,Choi15}.  

Work by \citet{Lodato06,Lodato07} has shown that low-angular momentum halos can provide favorable conditions for direct collapse; low-mass halos with large spin are less likely to form direct collapse black holes.  While some authors choose to require direct collapse occur only in low-spin halos \citep[e.g.][]{Volonteri10,Agarwal13}, we do not apply such a criterion because it is not consistent with our recipe for MBH formation based solely of local gas physics.  Furthermore, the available angular momentum transport processes would occur below our resolution limit \citep{Choi13}.  Such a requirement has previously been excluded in other work \citep[e.g.][]{Bonoli14} based on the justification that the dark matter and gas angular momentum are decoupled at scales below the virial radius.
%%%%%%%%%%%%%%%%%%%%%%%%%%%%%%%%%%%%%%%%%%%%%%%%%%%%%%%%%%%%%%%%%%%%%%%%
\subsection{Implementing the direct collapse model in \textsc{Gasoline}} \label{MBHcode}
%%%%%%%%%%%%%%%%%%%%%%%%%%%%%%%%%%%%%%%%%%%%%%%%%%%%%%%%%%%%%%%%%%%%%%%%
MBH seed formation in \textsc{Gasoline} is designed to emulate the existing star formation physics in the code.  The star formation module allows cold, dense gas particles undergoing local gravitational collapse ($\nabla \cdot v < 0$) to form stars stochastically, following a probabilistic function that depends on the molecular hydrogen abundance and the dynamical formation time of the particle \citep{Christensen12}.  The gas density must exceed the threshold density for star formation of 0.1 cm$^{-3}$ and must be colder than $10^4$ K, however, work by \citep{Christensen12} demonstrated that a star formation recipe driven by the H$_2$ fraction of the gas provides a more realistic spatial distribution of star formation in spiral galaxies.  The dependence of the star formation efficiency on the H$_2$ fraction supersedes the density and temperature thresholds by preferentially forming stars in regions abundant in molecular Hydrogen.  If a gas particle meets all of the relevant criteria, the probability that it will form a star is given by
\begin{equation}
P=\frac{m_{gas}}{m_{star}}\left(1-e^{c^{*}X_{\rm H_2}\Delta t/t_{form}}\right)
\end{equation}
in which the prefactor is the ratio of gas particle to initial star particle mass, $c^*$ is a free parameter set to 0.1 to recreate observable relations including the Kennicut-Schmidt law and Tully-Fischer law \citep{Governato10}, $t_{form}$ is the dynamical time for the gas particle, and $\Delta t$ is the interval between star formation episodes, set to 1 Myr.

Similarly, we allow MBH seeds to form stochastically from gas particles in environments that meet the convergence and temperature requirements for star formation. Gas particles considered for black hole candidacy must be $100$ times more dense than star forming gas ($10$ cm$^{-3}$), metal free to a tolerance of $Z<10^{-6}$, and experience Lyman-Werner radiation above a tuneable threshold that we specify at runtime, $J_{\rm crit}$.  If a gas particle meets the minimum MBH formation criteria, it is assigned a formation probability modeled after that used for star formation:
\begin{equation}
P_{BH}=1-e^{\frac{-J}{J_{\rm crit}}\Delta t/t_{form}}
\end{equation}
in which $J$ is the local Lyman-Werner flux level and $J_{\rm crit}$ is the critical Lyman-Werner radiation threshold employed in the simulation.

The Lyman-Werner radiation from stars in \textsc{Gasoline} is modeled as a function of stellar mass and age, as described in detail in \citet{Christensen12}.  The Lyman-Werner stellar luminosity is calculated as a function of the age and mass of the star particle via STARBURST99 \citep{Leitherer99} synthesis models for starforming regions assuming a \citet{Kroupa93} initial mass function.  The Lyman-Werner flux then propagates using the gravitational tree structure to approximate locality and assuming optically thin gas.  \citet{Christensen12} tested the accuracy of the Lyman-Werner model with a set of isolated disk galaxies from which they found that the computed Lyman-Werner flux value is within a factor of ten of the theoretical Lyman-Werner flux value in the optically thin limit.

The \textsc{Gasoline} black hole formation module allows MBH seeds to form without regard to global halo properties or the gas particle's position in the halo; rather, only the local gas properties dictate MBH seed formation.  This approach has the strong advantage that MBHs are not required to ``know'' anything about the large-scale properties of its parent halo.  MBH seeds are represented by sink particles, and form at a mass of $2.7\times 10^4 M_{\odot}$, taking on the entire mass of the parent gas particle.

Black holes in \textsc{Gasoline} are allowed to grow via isotropic gas accretion according to the Bondi-Hoyle formalism \citep{Bellovary10}.  Gas accretion is capped at the Eddington limit, assuming 10\% radiative efficiency.  A fraction of the rest-mass energy of the accreted gas is converted to thermal energy, which is then isotropically distributed to the surrounding gas \citep{DiMatteo05}.  This energy dissipates according to a blast-wave feedback approach similar to the supernova feedback recipe of \citet{Stinson06}, in that each affected particle's cooling is disabled for the duration of that particle's accretion time step.  MBH seeds can also grow through MBH-MBH mergers, which occur when the separation between two sink particles is less than two softening lengths and  their relative velocities are small: $\frac{1}{2} \Delta \vec{v}^2 < \Delta \vec{a} \cdot \Delta \vec{r}$, where $\Delta \vec{v}$ and $\Delta \vec{a}$ are the differences in velocity and acceleration of the two black holes, and $\Delta \vec{r}$ is their separation.

While our MBH seed model does not proscribe multiplicity of MBH seeds in a single halo, we include a mechanism to prevent spurious MBH overproduction.  We allow multiple MBH seeds to form at the same time step only if the particles are separated by more than the black hole merger separation criterion of two softening lengths.  If more than one gas particle meets the MBH formation criteria within this volume, only the most bound particle becomes an MBH seed, and the remaining MBH candidates revert to their parent gas particles.  Since black hole accretion and feedback occur on the smallest timescales in the simulation, this code modification gives the MBHs an opportunity to modify their environments via accretion feedback before the next episode of star and black hole formation, thus preventing the formation of multiple black hole particles in one place at effectively the same time.

%%%%%%%%%%%%%%%%%%%%%%%%%%%%%%%%%
\section{Results} \label{Results}
%%%%%%%%%%%%%%%%%%%%%%%%%%%%%%%%%%%%%%%%%%%%%%
\subsection{Massive black hole seed formation}
%%%%%%%%%%%%%%%%%%%%%%%%%%%%%%%%%%%%%%%%%%%%%%
\begin{table*}[t]
\begin{center}
\begin{tabular}{lccccccc} 
 \hline
 Name & $J_{\rm crit}$ & $N_{seed}$ & \thead{$N_{BH}$  \\ at $z=5$} & \thead{max($M_{BH}/M_{\odot}$)\\ at $z=5$} & $z_{first}$ & $z_{last}$ \\ 
\hline
 $J1000$ & $10^3 J_{21}$ & 67  & 15 & $2.7\times 10^5$ & 28 & 9.7 \\
 $J300$ & $300 J_{21}$  & 124 & 24 & $5.9\times 10^5$ & 28 & 8.7 \\
 $J30$ & $30 J_{21}$   & 849 & 88 & $3.1\times 10^6$ & 24 & 5 \\
 \hline
\end{tabular}
\caption{\small{\textsc{simulation parameters} Summary of selected results of the simulations presented in this paper. (1) Simulation name, (2) $J_{\rm crit}$ in units of $J_{21}$ ($10^{-21}$ erg s$^{-1}$ cm$^{-2}$ $sr^{-1}$ Hz$^{-1}$), (3) Total number of MBH seeds formed, (4) Number of black holes at $z=5$, (5) Mass of the most massive black hole at $z=5$, (6) Redshift of the first MBH seed formation event, (7) Redshift of the final MBH seed formation event.}}
\label{SimSummary}
\end{center}
\end{table*}

Table~\ref{SimSummary} summarizes our simulations and characterizes the number, maximum mass, and formation epoch of MBH seeds.  Note the vast difference between the $J30$ results and the higher Lyman-Werner thresholds.  A lower $J_{\rm crit}$  allows more MBH seeds to form, allows seeds to form at lower redshifts, and increases the final mass of the most massive MBH by redshift 5.  While this trend holds true in all runs, the difference between the $J300$ and $J30$ simulations is extreme.

\begin{figure}[ht]
  \centering
	\includegraphics[width=\columnwidth]{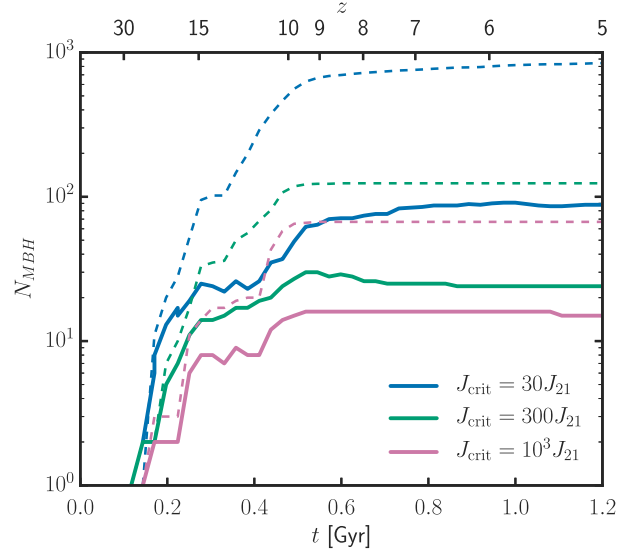}
	\caption{\small{\textsc{census of mbh formation} Solid lines show the number of MBHs present in each simulation over time.  The dashed lines show the total number of MBH seeds that have ever formed in each simulation.  MBH-MBH mergers account for the discrepancies between the total number of MBH seeds that have ever formed by a specific redshift and the current number of MBHs at that time.}}
	\label{SinkHistory}
\end{figure}

Figure~\ref{SinkHistory} takes a census of MBHs in each experiment.  Regardless of $J_{\rm crit}$, MBH seeds begin to form within the first $0.2$ Gyr.  Solid lines show the number of MBHs present in each simulation, whereas the dashed lines show the total number of MBH seeds that have ever formed in each simulation.  By comparing the solid and dashed lines, we can see a rapid series of MBH mergers between z$\sim 15-10$. This active merger epoch takes place before reionization, making it unlikely to observe using electromagnetic messengers but a promising source of gravitational waves detectable by LISA \citep{LISA17}.  MBH formation diminishes rapidly after $z\sim9$ in all simulations, but decreasing $J_{\rm crit}$ prolongs the epoch of seed formation.

\begin{figure}[ht]
  \centering
	\includegraphics[width=\columnwidth]{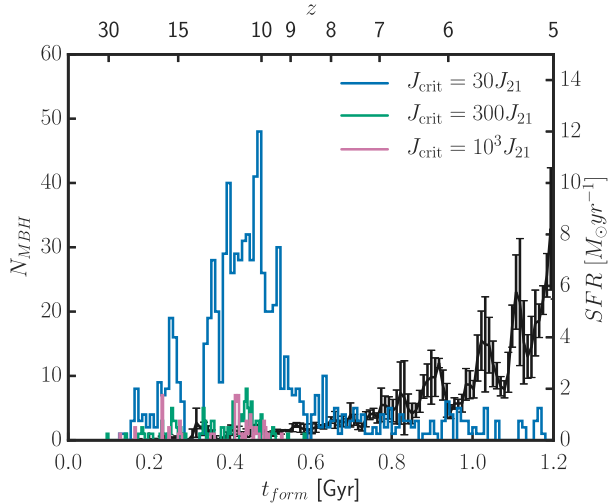}
	\caption{\small{\textsc{mbh formation rates}  The mean star formation history of the three simulations is shown in black with one-sigma error bars to demonstrate the variation in star formation rate between the three simulations. The MBH formation rate of the three simulations shows that lower $J_{\rm crit}$ values allow not only more MBH seeds to form, but additionally that a low $J_{\rm crit}$ threshold allows MBH seeds to form after MBH formation has dropped off with larger $J_{\rm crit}$ thresholds.  At high redshift, $z\gtrsim 15$, MBH formation occurs in sporadic bursts.  The rate of MBH formation peaks in all three simulations before $z=10$.  Only in the $J30$ simulation, with the most lenient MBH formation criteria, do MBH seeds form later than $z=9$, and may continue after $z=5$.}}
	\label{SinkFormationHistory}
\end{figure}

 Figure~\ref{SinkFormationHistory} compares the redshift evolution of black hole formation and star formation in the three simulations.  Though the number of MBH seeds differs, all three runs feature a peak in MBH seed formation near $z=12-10$ and rapidly drops off thereafter.  Though the $J30$ simulation continues to make a few new MBH seeds at lower redshifts, by $z=5$ most halos are sufficiently polluted with metals to quench MBH seed formation \citep{Bellovary11}. Still, we observe some seeds still forming at the end of our simulations, hinting at the possibility of a rare population of MBH seeds forming at low redshit.

\begin{figure}[ht]
  \centering
	\includegraphics[width=\columnwidth]{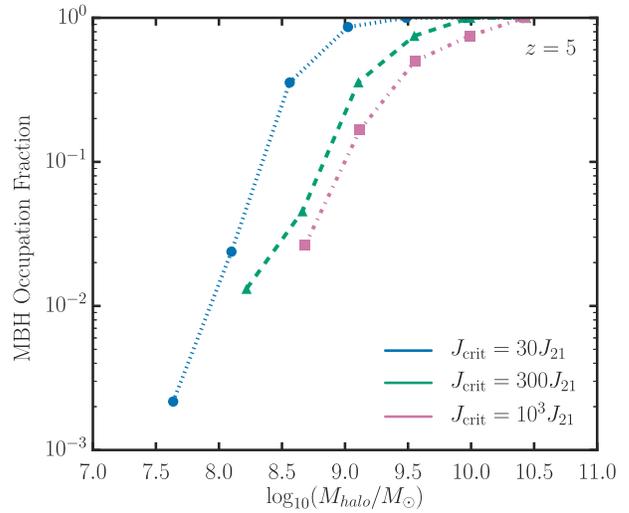}
	\caption{\small{\textsc{mbh occupation fractions} A comparison of the MBH halo occupation fractions from simulations with three different $J_{\rm crit}$ values shows that lower $J_{\rm crit}$ thresholds allow MBH seeds to form in lower-mass halos, and that low-mass halos are more likely to host MBHs.  At $z=5$, a minimum of half of all halos more massive than $3\times 10^9 M_{\odot}$ host an MBH seed, regardless of the $J_{crit}$ threshold.}}
	\label{OccupationFraction}
\end{figure}

The occupation fraction of massive black holes, particularly in local dwarf galaxies \citep{Miller15}, likely encodes information about the MBH formation process.  This metric may provide clues as to the initial MBH seed mass \citep{vanWassenhove10}, the MBH formation efficiency \citep{Volonteri08}, merger history \citep{Menou01}, and the relative contribution of early and late MBH formation epochs \citep{Tanaka09}.  Unfortunately, the high redshift occupation fraction is highly uncertain \citep{Lippai09}.  We define the MBH occupation fraction as the fraction of halos in a mass bin of width $\log_{10}M/M_{\odot}=0.5$ that host at least one MBH.  In Figure~\ref{OccupationFraction} we investigate the effect of varying $J_{crit}$ thresholds on the MBH occupation fraction at the end of our simulations.  Even under the most stringent formation criteria, a minimum of 50\% of halos with masses larger than $3\times 10^9 M_{\odot}$ at $z=5$ host at least one MBH.  These halos are expected to grow to be highly massive ($10^{11}-10^{12}M_{\odot}$) by the present day \citep{McBride09}.  The difference between these thresholds is most apparent below $\sim 3\times 10^8 M_{\odot}$, arguing that the occupation fraction of Milky Way-mass disk galaxies at redshift zero may inform $J_{\rm crit}$.  Mapping local occupation fraction to high redshift is more complicated in general but far more so for low mass galaxies, where gravitational wave recoil and alternative seed formation channels may become increasingly important \citep{Schnittman07,Blecha08,HolleyBockelmann08,HolleyBockelmann10}.  Incorporating these effects in a statistical sense may require a close collaboration between state-of-the-art simulations like these and semi-analytic techniques.

%%%%%%%%%%%%%%%%%%%%%%%%%%%%%%%%%%%%%%%%%%%%%%
\subsection{Sources of Lyman-Werner radiation}
%%%%%%%%%%%%%%%%%%%%%%%%%%%%%%%%%%%%%%%%%%%%%%
Primordial star formation is thought to provide the essential Lyman-Werner flux to spur direct collapse black hole formation, but the proximity to these star formation sites is not well-constrained.  One favored model of isothermal direct collapse invokes the synchronized growth of a pair of neighboring atomic cooling halos \citep{Dijkstra08,Visbal14b}.  As one of the halos crosses the atomic cooling threshold, it experiences a burst of star formation that irradiates the secondary neighbor.  If the Lyman-Werner radiation generated by this stellar population is sufficient, it prevents molecular hydrogen cooling in the secondary halo, setting up the right conditions for direct collapse.

\begin{figure}[ht]
  \centering
	\includegraphics[width=\columnwidth]{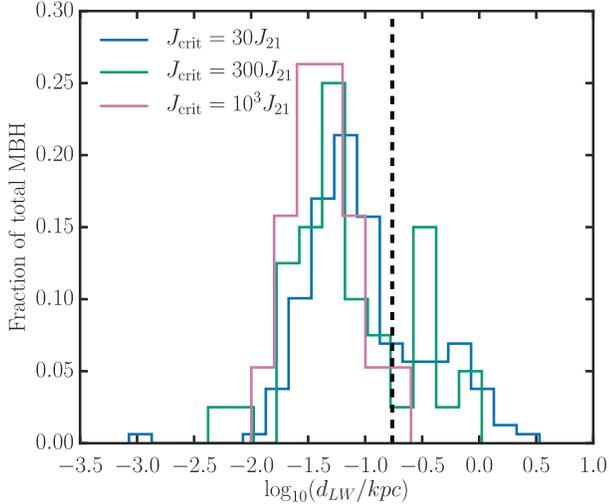}
	\caption{\small{\textsc{mbh-lw source separations}  Distributions of distances between MBH formation sites and the estimated dominant source of LW radiation.  The dashed vertical line marks one softening length.  For most MBHs, the nearest source of LW radiation is located inside the host halo and less than one softening length from the MBH formation site.}}
	\label{d2LW}
\end{figure}

We investigate the proximity and synchronization of Lyman-Werner sources and direct collapse sites to characterize the types of scenarios that favor direct collapse.  Since the strength of Lyman-Werner flux scales as $\sim d^{-2}$, we approximate the dominant source of Lyman-Werner radiation for each MBH seed as the nearest star forming region that formed prior to seed formation.  To demonstrate the validity of this assumption, we provide some examples here.  At $z=24$ in the $J30$ simulation, an MBH forms in a $2\times 10^7 M_{\odot}$ halo.  The dominant source of Lyman-Werner radiation is a star particle that formed in this halo less than $1$ Myr before the MBH seed and $0.5$ kpc from the direct collapse site.  The flux from this starforming region is six orders of magnitude larger than the flux from the only other starforming region at this time, which is $\sim 50$ kpc away, and has triggered direct collapse in its own halo.

As the simulation evolves and star formation increases, more distant starforming regions begin to contribute a more significant fraction of the total Lyman-Werner flux on the direct collapse sites.  At $z=10$ in the $J30$ simulation, an MBH forms in a $2\times 10^8 M_{\odot}$ halo that previously formed one star particle approximately $1$ kpc from the direct collapse site.  The Lyman-Werner flux from this single star particle comprises $\sim 75\%$ of $J_{\rm crit}$ while the remainder of the supplemental flux comes from a neighboring starforming halo $\sim 75$ kpc away, bringing the total Lyman-Werner flux at the direct collapse site above $J_{\rm crit}$.

In Figure \ref{d2LW}, we show the distance from MBH seed formation sites to the nearest star particle.  We find that $\gtrsim 80\%$ of the irradiating sources are inside the same halo and less than $2$ kpc away from the radiation source.  MBH seeds in these simulations form in pockets of low-metallicity gas that are exposed to Lyman-Werner radiation from star forming regions within their host halos.  With lower $J_{\rm crit}$ thresholds, MBH seeds are able to form at larger distances from Lyman-Werner sources and farther from the centers of the host halos.

We do notice synchronization between star and MBH seed formation within the same halo; regardless of $J_{\rm crit}$, the majority of Lyman-Werner sources formed less than 5 Myr before the collapse of their respective MBH pairs.  This contemporaneity of MBH seeds and nearby stars hints at a tendency for Lyman-Werner sources and MBHs to form in synchronized pairs, as predicted by \citet{Dijkstra08,Visbal14b}, although the pairs that we find in these simulations tend to form in a single halo.  In this scenario, stars form and a MBH seed forms shortly afterwards before metals can contaminate the halo.  
\begin{figure*}[ht]
	\centering
	  \subfloat{\includegraphics[width=\columnwidth]{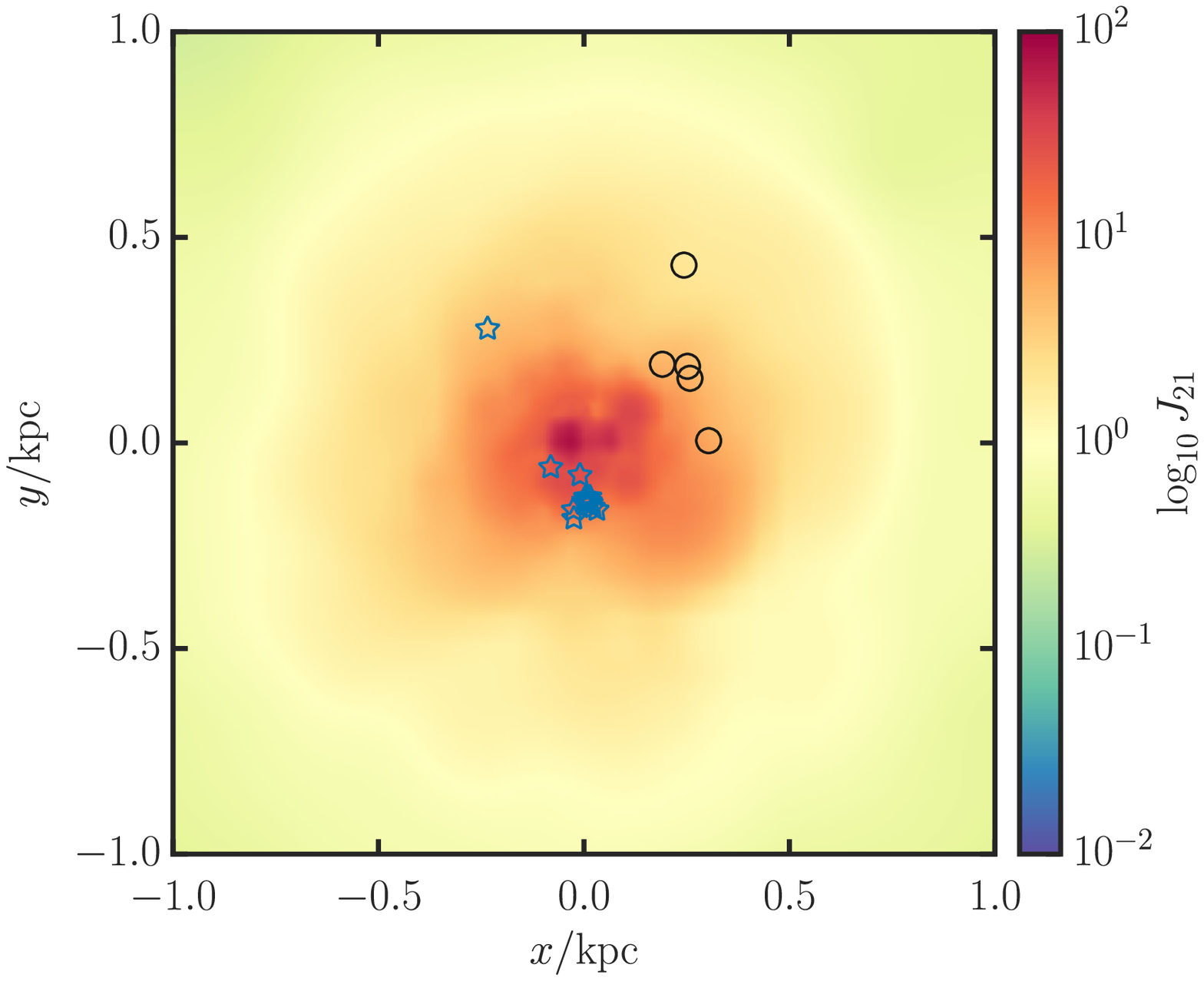}}
	  \subfloat{\includegraphics[width=\columnwidth]{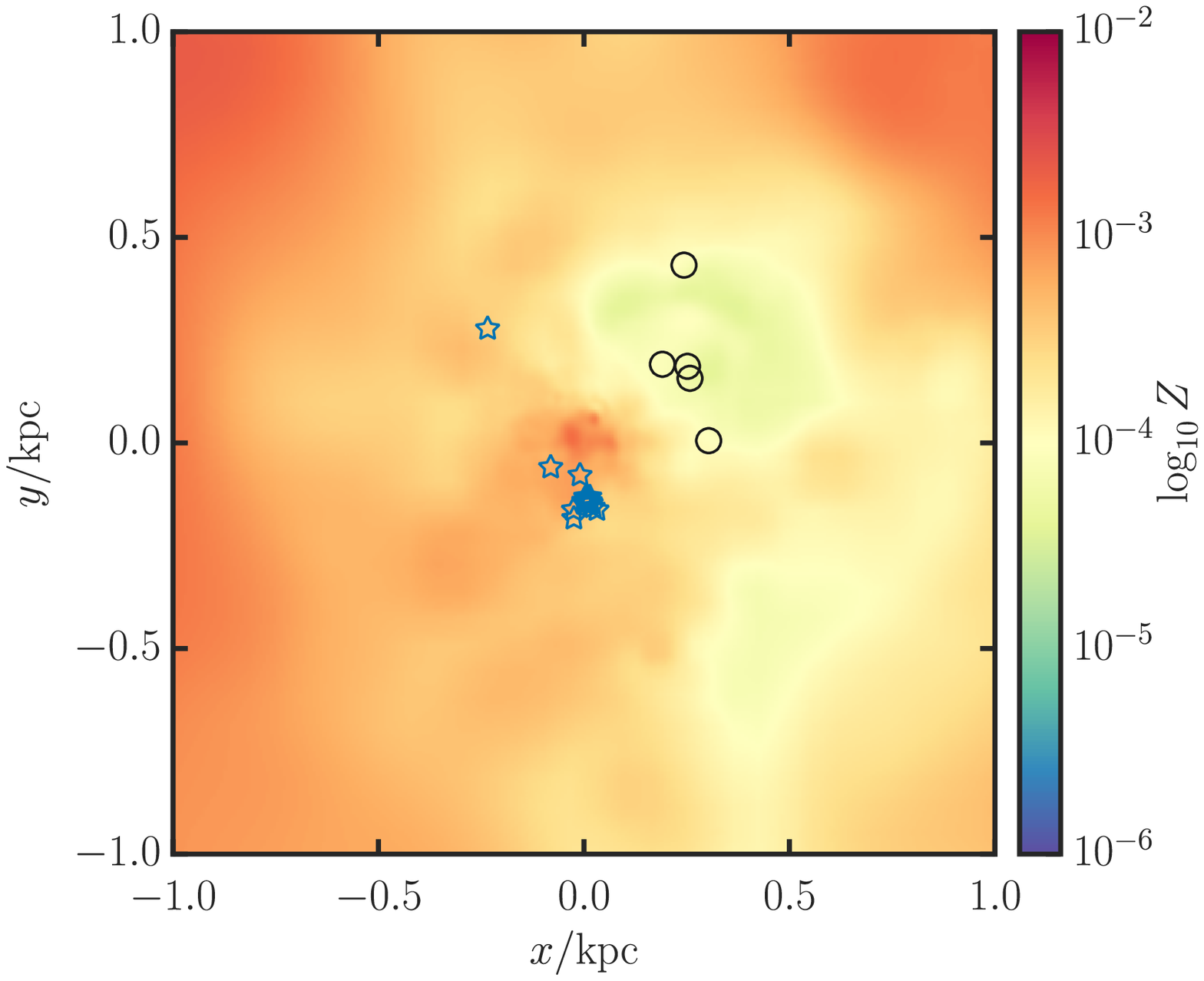}}
	  \caption{\small{\textsc{an atomic cooling halo pair candidate}} Left: Massive black hole seed positions are marked by black circles in this $z=13$ line-of-sight averaged Lyman Werner flux image of the atomic cooling halo pair candidate in the $J30$ simulation.  The five MBH seeds are members of the $2.5\times 10^8 M_{\odot}$ subhalo that has not formed stars.  Blue stars mark the positions of nearby star particles, which are members of the $8\times 10^8 M_{\odot}$ parent halo.  Right: MBH and star particle positions are marked in a line-of-sight averaged metallicity image of the atomic cooling halo pair candidate.  MBH seeds form in regions that experience high Lyman-Werner flux but are still sufficiently metal poor.}
	\label{haloPair}
\end{figure*}
In the $J30$ simulation, we observe a single instance of seed formation that closely resembles the atomic cooling halo pair scenario, which we illustrate in Figure~\ref{haloPair}.  This scenario is quite similar to that proposed in \citet{Visbal14b}.  At $z=13$, a halo of mass $2.5\times 10^8 M_{\odot}$ crosses the virial radius and becomes a subhalo of a more massive halo ($8\times 10^8 M_{\odot}$).  While the subhalo remains pristine, the more massive parent halo begins forming stars, providing enough Lyman-Werner radiation to form five MBH seeds in the subhalo.  From this one example, we conclude that while atomic cooling pairs are a viable seed formation mechanism in general, it is too rare to comprise the bulk of the seed formation in our models.  Alternatively, we show that direct collapse black holes can, and predominantly do, form in low-metallicity pockets of halos with prior star formation.
%%%%%%%%%%%%%%%%%%%%%%%%%%%%%%%%%
\subsection{Characteristics of halos that host MBH seeds}
%%%%%%%%%%%%%%%%%%%%%%%%%%%%%%%%%
The canonical picture of direct collapse allows a single MBH seed to form in a globally pristine halo with no history of star formation.  In contrast, our simulations indicate that this process is significantly more complex and should be treated as such.  This complexity could be expected by considering the more complicated way realistic halos form and evolve.  Even immediately after a halo forms, it is not spherically symmetric or homogeneous, but exhibits irregular morphology.  Likewise, gas collapse is not monolithic, but rather clumpy and stochastic.  Conditions inside these dark matter halos can be ripe for both star formation and direct collapse within a single halo, and may produce multiple MBH seeds.  In this set of simulations, it is not uncommon for MBHs to form in groups that quickly merge to form a more massive MBH.  Since we have taken precautions to prevent spurious MBH formation, as discussed in Section~\ref{MBH}, we believe the formation of multiple MBH seeds per halo is a real result, and not a numerical effect.

\begin{figure}[ht]
	\centering
	  \includegraphics[width=\columnwidth]{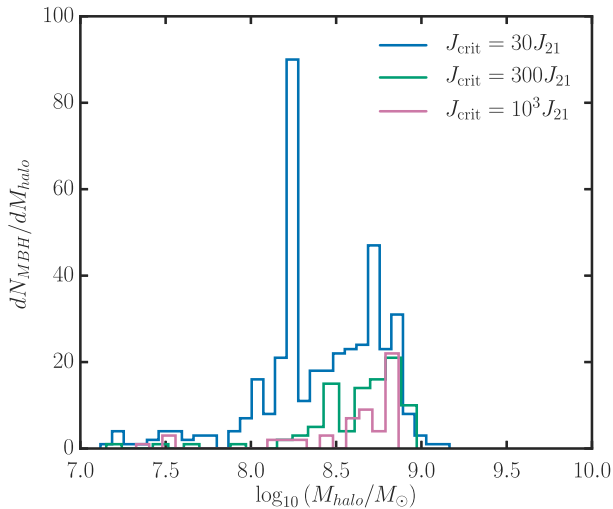}
	  \caption{\small{\textsc{masses of mbh-forming halos} A distribution of halo masses at the time of MBH seed formation in each simulation show that MBHs form preferentially in halos with masses greater than $10^8 M_{\odot}$.}}
		\label{dNdM}
\end{figure}

\begin{figure}[ht]
	\centering
	  \includegraphics[width=\columnwidth]{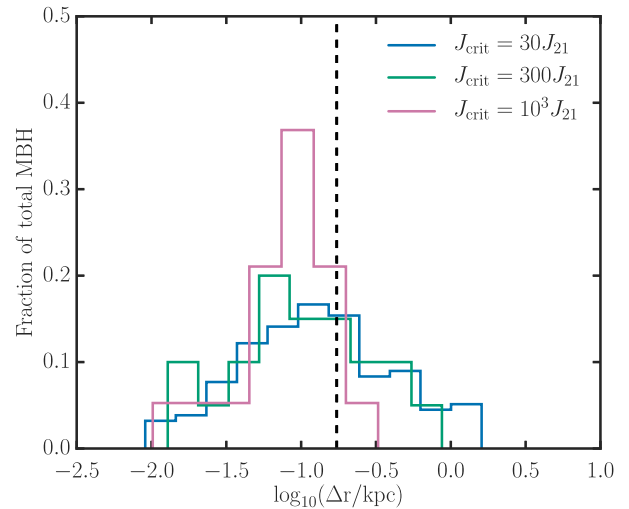}
	  \caption{\small{\textsc{mbh-host halo center separations}  Distributions of offsets between MBH formation sites and host halo centers of mass.  The dashed vertical line marks one softening length.  Most MBHs form within one softening length of the host halo's center of mass.}}
	  \label{d2center}
\end{figure}
In Figure~\ref{dNdM} we show the masses of seed host halos at the time of seed black hole formation in our three simulations.  While we do not explicitly constrain the halo masses of potential seed hosts, the minimum gas density threshold we set effectively acts as a proxy for preferential MBH formation in more massive halos.  Similarly, while there are no explicit restrictions placed on where in the halo an MBH can form, the densities required to form an MBH create a natural tendency for direct collapse to occur at the centers of the host halos.  In Figure~\ref{d2center}, we show the distributions of offsets between the MBH formation site and the host halo center-of-mass.  The majority of MBHs form within one softening length of the center of the host halo.

In a comparison of global halo characteristics of halos that host MBHs and those that do not, we find a slight decrease in the mean halo concentration of MBH-hosting halos as compared to the mean halo concentration of all halos.  While this trend persists at all redshifts, it is not statistically significant. 

%%%%%%%%%%%%%%%%%%%%%%%%%%%%%%%%%
\section{Summary} \label{Summary}
%%%%%%%%%%%%%%%%%%%%%%%%%%%%%%%%%
In this work, we used cosmological simulations to study the formation of direct collapse MBH seeds under the influence of of spatially and temporally varying anisotropic Lyman-Werner radiation field and characterize the demographics of MBH formation sites.  In future work, we will examine MBH seed formation and early growth in a larger suite of simulations spanning $z=0$ halo masses from dwarfs to massive elliptical galaxies.

Building on the findings of \citet{Bellovary11}, that MBH seeds can form in less massive halos, we additionally find that direct collapse MBH seeds can form in halos with star formation history and non-zero metallicity.  Lower $J_{\rm crit}$ values allow MBH seeds to form more abundantly, at lower redshifts, and in less massive halos.  The sources of $H_2$ dissociating Lyman-Werner photons are likely to form in the same halo as the seed.  We have shown that, contrary to the canonical model, MBHs do not need to form singly in completely pristine halos.  This work indicates that conditions favorable for MBH formation may necessitate some previous star formation in the host halo, thereby causing MBHs to form in metal-enriched halos.  Multiple MBHs can, and do, form in a single halo, and these black holes often merge into larger seeds.

The limitations of this work stem from the challenge of incorporating physical processes that occur across large dynamic ranges in cosmological simulation.  The occupation fractions presented in this paper are upper limits because we do not account for gravitational recoil, which will cause some fraction of MBHs to evacuate their halos when merging.  We are not able fully resolve the collapse of gas into a MBH seed, which occurs on sub-parsec scales, and instead must represent this unresolved physics with sub-grid models that take as input the conditions of the gas physics at distances of $\sim 100$s of parsecs.  Another source of uncertainty is the Lyman-Werner escape fraction, which depends of the form of propogation of the ionization front \citep{Schauer17b}.  To correctly compute exact Lyman-Werner fluxes would require a full treatment of radiative transfer.  These simulations also do not account for X-ray radiation, which alters the gas chemistry by increasing the free electron fraction, thereby inciting H$_2$ formation \citep{Haiman96,Haiman01,Tanaka12}.

There are currently very few observational constraints on MBH formation, high-redshift occupation fractions, or MBH-MBH merger rates.  Future electromagnetic and gravitational wave observatories, including  {\sc JWST}, {\sc LISA}, {\sc Athena}, {\sc LUVOIR} and {\sc Lynx}, may offer some of the first chances to observationally constrain MBH formation scenarios.  {\sc LISA}, specifically, is the only mission that will be able to observe MBH-MBH mergers out to $z\sim20$, and so will uniquely observe the seed formation epoch directly \citep{LISA17}. We can track the expected gravitational strain amplitude of each MBH-MBH merger to predict the total stochastic gravitational wave signal from the population of MBH-MBH mergers as a function of time~\citep[e.g.][]{HolleyBockelmann10}; this will provide a foundation for {\sc LISA} observations to constrain MBH formation.  In a future paper, we will calculate the gravitational wave signal from early MBH seed mergers and discuss the potential detectability of these events with {\sc LISA}.

\acknowledgements
The simulations used in this work were run on XSEDE/TACC Stampede and Stampede2 under allocation TG-AST150063, NASA/Pleiades, and the ACCRE cluster at Vanderbilt University.  We used the following software packages in post-processing analysis: {\sc Pynbody} \citep{Pontzen13}, {\sc Matplotlib} \citep{Hunter07}, {\sc NumPy} \citep{Oliphant06}, {\sc PANDAS} \citep{McKinney10}, and {\sc Python} \citep{vanRossum95}.  JMB was supported by PSC-CUNY grant 60303-00 48.

\bibliography{LWpaper}

\end{document}